# Enhanced Data Integration for LabVIEW Laboratory Systems

Adriana Olteanu, Grigore Stamatescu, Anca Daniela Ioniţă, *IEEE Senior Member*, Valentin Sgârciu
University POLITEHNICA of Bucharest
adriana.olteanu@aii.pub.ro, gstamatescu@aii.pub.ro, anca.ionita@aii.pub.ro, vsgarciu@aii.pub.ro

*Abstract-* Integrating data is a basic concern in many accredited laboratories that perform a large variety of measurements. However, the present working style in engineering faculties does not focus much on this aspect. To deal with this challenge, we developed an educational platform that allows characterization of acquisition ensembles, generation of Web pages for lessons, as well as transformation of measured data and storage in a common format. As generally we had to develop individual parsers for each instrument, we also added the possibility to integrate the LabVIEW workbench, often used for rapid development of applications in electrical engineering and automatic control. This paper describes how we configure the platform for specific equipment, i.e. how we model it, how we create the learning material and how we integrate the results in a central database. It also introduces a case study for collecting data from a thermocouple-based acquisition system based on LabVIEW, used by students for a laboratory of measurement technologies and transducers.

## I. Introduction

Electrical engineering foundations have been studied by all the students of technical universities during the first two academic years, because many engineering domains use electrical devices of all kinds. However, the integration in the European educational system and the current society trends towards direct participation of many stakeholders require new methods of teaching, new software environments and new kinds of student-teacher partnerships [1]. The work in faculty laboratories is confronted with the following challenges:
- There is a large range of instruments used in measurements, even inside a single laboratory.
- Several instruments may be aggregated in ensembles, together with the measured subject and the acquisition system.
- Data have to be available in a common format that can be further processed, irrespective of the source they come from.
- Students should be able to access experimental results obtained with complex equipment that may not be available to them directly.

The solutions that are expected concern: i) adopting a simple process of acquisition, ii) supporting virtual access to procedures and data, iii) allowing measurements from a large variety of non-homogeneous instruments; iv) exporting data to other systems, for processing or for integrating with data originated from other sources. All these concerns may be attained with LabVIEW (Laboratory Virtual Instrument Engineering Workbench) [2] - the graphical development environment proposed by National Instruments. It has become widespread among students, professors and researchers alike, because it provides a high level of abstraction for further technical software development. The workbench supports a wide range of instruments, data acquisition devices and real-time controllers, with many input and output lines. It allows the developer to invest more in processing, instead on focusing on low level programming tasks for acquisition. For example, instead of several tens of text-based code lines for configuring the devices and actuators of a control loop, the developer places pre-defined graphical blocks in the block diagram, and s(he) interconnects them by means of wires, inside a while or for loop.

Important application scenarios are simulation of virtual laboratories, and remote access to real industrial equipment. LabVIEW was used both in education and in advanced research, in a large variety of fields [3]: feedback control theory for distillation columns, water flow / level, generator voltage / temperature, nuclear instrumentation, analogue electronics, radioactive disintegration, RC and RL circuits.

Our current aim is to combine the integrative power of LabVIEW with the EquiLAB educational platform [4], which we previously developed for laboratories based on non-homogeneous acquisition systems, including simple instruments, sophisticated equipment, sensor networks, and various hybrid ensembles.

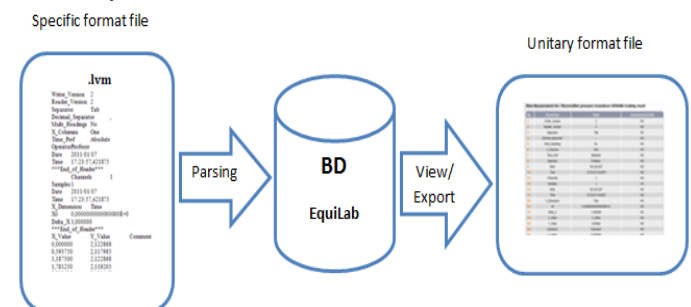

Fig. 1. The transformation of a lvm file in a unitary format.

The data files obtained from LabVIEW are parsed and stored in a central database, where they are saved in a unitary format. After that, authenticated users can view and / or export data to other processing software, irrespective of the original source of the measurements (see Fig. 1).

The paper describes the way we integrate a major workbench like LabVIEW with a system developed in-house, EquiLAB, and analyzes the particular case of a thermocouple-based laboratory ensemble, used for students' practical work. Chapter II presents the measurement ensemble and the necessary LabVIEW settings; chapter III describes the configuration of EquiLAB for this ensemble, and chapter IV presents technical details regarding data integration.

II. THE THERMOCOUPLE LABORATORY SYSTEM BASED ON LABVIEW

*A. The SYTHERM Ensemble*

The SYTHERM ensemble is part of the Intelligent Measurement Technologies and Transducers Laboratory at the Faculty of Automatic Control and Computers, University POLITEHNICA of Bucharest. It mainly serves a didactic purpose for the third year students, supporting practical applications of the Transducers and Measurement Systems course. The particular goal is to prove the utility of dedicated data acquisition modules, which capture signals directly from generator sensitive elements (i.e. J- and K-type thermocouples). One uses the processing, display and data logging functions of the LabVIEW development environment.

In order to acquire data and study the thermocouple characteristics corresponding to the requirements, one employs the laboratory set-up presented in Fig. 2. It contains the following components:

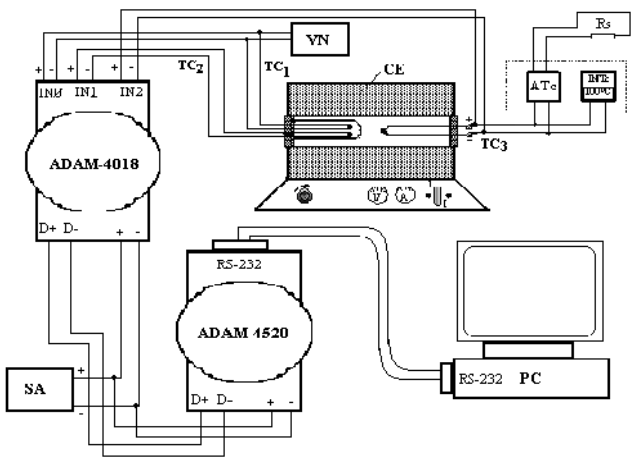

Fig.2 . Laboratory set-up for thermocouple signal acquisition using ADAM-4018 module [5].

- CE – Electric oven for thermocouple calibration;
- SA – Power supply 24VDC, 1A;
- VN – Digital voltmeter with 3 ½ digits;
- $TC_1$, $TC_2$ – J-type thermocouples, installed in a common enclosure;
- $TC_3$ – K-type thermocouple with no protective enclosure;
- ATc/INTc – Adapter for the K-type thermocouple, with local temperature display in $^oC$; because the adaptor ATc provides a unified continuous current output, load resistance $R_s$ is employed with a typical $250\Omega$ value (as this output current is not subsequently measured, the current output can be short-circuited);
- ADAM-4018 – Thermocouple signal acquisition module; it can acquire signals simultaneously from 8 thermocouples, with the mention that the last two inputs have to use a common ground (details can be found in the datasheet);
- ADAM-4520 – RS-485 to RS-232 conversion module;
- PC – Personal computer with dedicated LabVIEW software application for analog thermocouple signal acquisition, using the ADAM-4018 module through dedicated drivers [5].

The purpose is to build the experimental set-up, by connecting the transducers to the interface modules and then to the PC. It is necessary to define an appropriate configuration, consisting of scale and sensitivity adjustments, serial port settings, and software structure for a virtual instrument. After that, the application is run in order to acquire data from the temperature process. Data is saved in either Excel (*.xls*) or LabVIEW (*.lvm*) measurement file formats. With these values, students go on to compute the non-linearity error/static characteristic. This is performed off-line, based on the acquired data, the reference voltage values and the local indicator display. The non-linearity error equation is (1)

$$\varepsilon_{nel_i}[\%] = \frac{\left|T_{real_i} - T_{ref_i}\right|}{T_{ref_{30}} - T_{ref_i}} \cdot 100, \quad (1)$$

where:
- $T_{real}$ is computed by subtracting the J-type thermocouple voltage from the real voltage output, at the reference ambient temperature;
- $T_{ref}$ is the reference temperature for the J-type thermocouple;
- $T_{ref30}$ is the reference temperature at the ambient temperature.

Thus, the linearization functionality and the reference junction temperature compensation of the ADAM-4018 module can be validated.

The dynamic characteristic of the J-type thermocouple is computed by modeling it as a first order delay system and deriving the time constant T graphically. The experimental procedure is performed by cooling the temperature and periodically sampling the output values, until a steady state regime is achieved (i.e. low variation of the temperature over several measurements).

*B. LabVIEW Acquisition*

LabVIEW can be used for creating a virtual instrument and building applications based on a dual approach: the development of the functionality and that of the user interface. There is a seamless integration with either built-in or third-party hardware. Moreover, applications can be easily distributed by creating installers that include the necessary run-time information for deploying on the target machine. We use two parts of LabVIEW programs:
- The *Front Panel* - the user interface consisting of passive numerical and graphical indicators, and active elements that require user input, such as: buttons, switches, text boxes, etc.
- The *Block Diagram* – interconnecting these elements with functions and various programming structures, in order to obtain the desired functionality.

The LabVIEW measurements are saved in the *.lvm* format [6] and are obtained with specific create and read blocks, e.g. Read/Write. The result is an ASCII text file with one-dimensional data, suitable for easy reading and parsing within data processing tools. It is designed for small and medium datasets, in contrast to large data collection applications with many channels and high sampling frequency, where binary formats are more suitable. This makes it best suited for our laboratory scenario. A *.lvm* file starts with a header section that can be configured by the user, and continues with segments that include the actual data. The header gives details that help data parsing, by including the separator character and the time format for stamping the data. Each segment can also have its own header. The *.lvm* logging is implemented in our application with dedicated blocks that handle incoming data from three measurement channels. We use *Set Dynamic Data Attributes* virtual instruments for labeling the three channels, as well as for time-stamping the data. A source code fragment is shown in Fig. 3 and the configuration window is shown in Fig. 4.

The resulting file structure contains a header, describing the measurement session, followed by the data - with one column per temperature channel. For each measurement, the timestamp is expressed in the form of an offset related to the absolute timestamp included in the header. An example of such a file is shown in Annex 1.

### III. CONFIGURING THE EDUCATIONAL PLATFORM FOR THE THERMOCOUPLE SYSTEM

*A. The EquiLAB platform*

The EquiLAB educational platform was developed to fulfill the following objectives:
- Modeling measurement systems / instruments using dedicated concepts;
- Providing unitary educational support for trainers and academics, based on Web 2.0 and e-learning technologies [7];
- Integrating data from complex apparatus, equipment, industrial or experimental ensembles.

Fig. 5 outlines the three components of our platform, corresponding to these objectives.

In the *Measurement System Modeling* module, new equipment can be modeled by filling in the following fields: name, producer, short description, webpage, picture, visual model, parsing procedure, and data file extensions. For modeling, we use specific concepts that describe what is measured, what results are obtained, and in what conditions. For this purpose we model each system by defining parameters that characterize the concepts below [8]:
- The Instrument Setup - realized before data acquisition;
- The Data - representing different physical properties amounts;

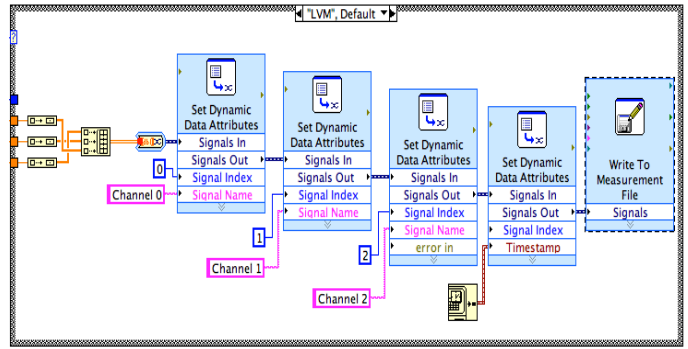

Fig. 3. Block Diagram Fragment with .lvm Logging.

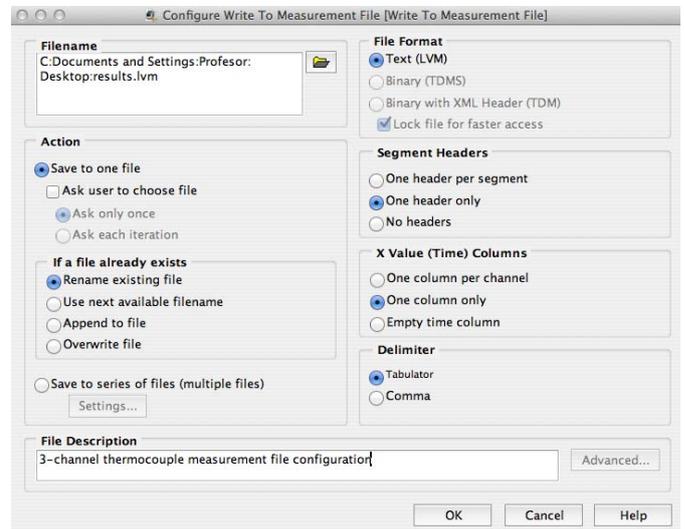

Fig. 4. The LabVIEW Configuration Window.

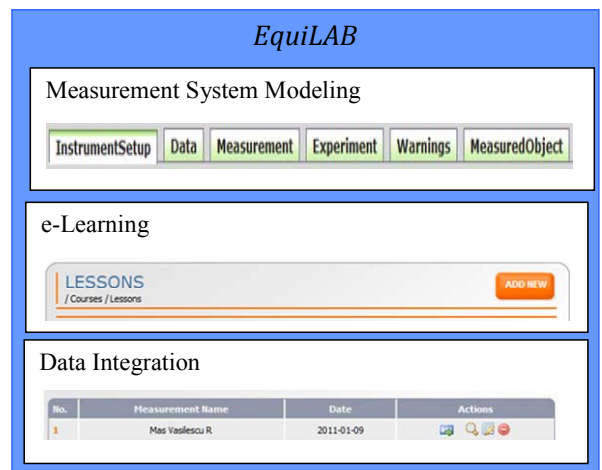

Fig.5 . The EquiLAB Educational Platform Modules.

- The Measurement information, e.g. the responsible person, or the date;

- The Experiment characterization – for being able to reproduce similar measurement conditions, constraints, and procedures;
- The Warnings – raised by the instruments that compose the system;
- - The Measured Object, i.e. a measurement ensemble or a simple material sample.

Each parameter has associated a measurement unit and a data type. One should specify the measurement units according to the International System of Units (e.g. Second, Radian, Tesla, Ampere, Celsius Degree) and should choose the parameter types (Integer, Real, Boolean, Time, Date, Enumeration or String). It is necessary to indicate whether the value must be imported from a file or read from the keyboard. This model definition is only required once, when integrating a new equipment.

The *e-Learning* module involves three types of actors: teachers, students and platform administrator. Each of these roles has specific characteristics. The teacher introduces and characterizes new instruments, manages the content of lessons, generates Web pages and assigns the corresponding equipment for the laboratory tasks [9]. The student visualizes the associated lessons and introduces measured data in the integrated database. The administrator creates accounts for teachers and students, manages the rights associated with these accounts and deletes users. The EquiLAB system accepts personal contributions and user generated content; it is also configurable, because teachers can add new criteria for equipment characterization.

The *Data Integration* module allows one to import data from a measurement system. The user selects the data file, guided by the program to associate the data to the appropriate file extension for that equipment; after that data are read, parsed and stored in the database. The parsing operation brings the information from a heterogeneous into a uniform format. The measurements are obtained in different types of files (e.g. *.msr*, *.txt*, *.lvm*, *.coi* etc.). Each of these files has a different data format. Therefore, when modeling new measurement equipment, the user has the opportunity to assign a specific parsing procedure and the type of files generated from measurements. Information stored in the database can then be queried according to the above classification criteria (data, experiment, instrument etc.).

The user can also export the data in a standard format, allowing further processing of information provided by the measurement equipment. For this purpose we chose two generally used targets:
- Excel - preferred for its complex capabilities of introducing formulas and representing charts, even if its basic functionalities are known even before entering the faculty) and
- XML (eXtensible Markup Language) – selected because it is a standard for interchanging data and it is used as a basis for program interoperation in many scenarios [10].

*B. Defining SYTHERM on the EquiLAB Platform*

SYTHERM was modeled according to the previously described structure and added to the EquiLab educational platform. It was associated the UPB Measurement Laboratory as producer, because this system was realized locally, as an educational environment. The valid extension was set to *.lvm*. Choosing the parsing procedure was an important step. As the LabVIEW parser was previously defined for other measurement ensembles, we just had to select an existing program during the configuration stage.

The parameters are mapped to the above presented concepts, i.e.:
- the Data category contains: *X_Value*, *Channel_0*;
- the Measurement category is characterized by: *Operator*, *Date*, and *Time*;
- the Experiment category is composed of: *Channels*, *Separator*, *Decimal_Separator*, *Multi_Headings*, *X_Columns*, *Time_Pref*, *X_Dimension*, *X0*, and *Delta_X*.

Several parameters from the *.lvm* file are ignored, because they are not relevant to the measurements, for example: *Writer_Version* and *Reader_Version*. The InstrumentSetup, MeasuredObject and Warnings parameters are not present in this case study.

IV. INTEGRATION OF STUDENTS' EXPERIMENTAL DATA

*A. Data Integration Workflow*

The integration of experimental data from heterogeneous equipment is based on the idea of storing the parameters and the values separately. The parameters are introduced in the equipment model, which is only defined at the beginning, using a dedicated model editor.

Fig. 6 presents the data integration process. The data import sets the parameter values each time one performs a new measurement with that equipment. The entries may come directly from a software system or from a laboratory operator. Experimental data are parsed and saved into a database. Integrated data may be used for visualization and for export in standard formats, but also for inter-operation with external programs for Computer Aided Design. For each measurement it is possible to execute the following actions: remove, edit, save, view, export to .xml, export to .xls.

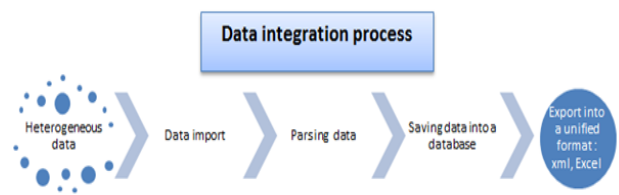

Fig. 6. Data integration process.

*B. Parsing Experimental Data*

Data from the measurement files are parsed and saved in the database organized in accordance with the concepts previously used for grouping parameters. Users can import measurements obtained in the laboratory. An instrument or a measurement ensemble may produce one or more types of files, each one associated to a parsing procedure, whose name contains the extension and the equipment name, as presented in Fig. 7.

When new equipment is integrated into the system, the user must choose one or more procedures for parsing and the extensions of the data files. For example, for the characterization of magnetic [11], several examples are:
- *txt* for Vibrating Sample Magnetometer;
- *mes*, *set* and *coi* for Hysterezisgraph;
- *hys*, *mes*, *msr* and *msa* for Single Sheet Tester (SST).

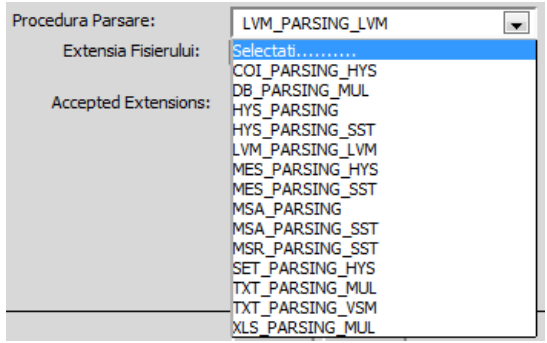

Fig. 7. The page for choosing parsing procedure

Each of these files has a particular content for each instrument. Thus, the Hysterezisgraph file with the *.mes* extension has another form and content than the *.mes* file for Single Sheet Tester. Multiple measuring instruments may have associated files with the same extension, but different formats and significance. In conclusion, the solution is to develop different procedures for parsing, in respect with the equipment name and the data file extension.

A special case is the *lvm* file extension, having the same format irrespective of the ensemble connected to LabVIEW. The *lvm* file content may be slightly different from one case to another, depending on the settings made with LabVIEW software, and on the person involved in the measurement process, but the format remains the same, so it may be interpreted taking into account the parameters introduced at modeling.

The parsing procedure is performed using the PHP open source technology. It consists of uploading the file obtained from the measurement and parsing it according to the designated program, which searches the values correspondent to the previously defined parameters. For each file type, we implemented a dedicated parsing function that translates data from the specific format to a single, integrated format, stored in a database.

*C. The Database Model*

The database was developed using MySQL and it contains data about the equipment, with parameters divided in the same groups as used for modeling. The database consists of 23 tables, divided in 5 interconnected categories (see Fig. 8).

The table names are suggestive, e.g. for *t_eqp_equipments* "t" represents the table, "eqp" corresponds to *eqp_number* - the primary key used in that [12] - and "equipments" shows that it stores information about the equipment. Besides the primary key, in the *t_eqp_equipments* table there are other important fields, shown in Fig. 9, together with the structure of two other tables: *t_psf_parsingfunction*, and *t_efe_equipment*.

Between the *t_eqp_equipments* and *t_psf_parsingfunction* tables there is a many-to-many relationship, because the equipment can allocate several parsing procedures, depending on the files obtained from measurements for the ensembles of instruments. In order to design the many-to-many relationship and to normalize the database, one introduced a link table: *t_efe_equipmentfileextension*, which contains *eqp_number* and *psf_number* - the key fields from the *t_psf_parsingfunction* and *t_eqp_equipments* tables. The parsing procedures are stored as *psf_name*, in the *t_psf_parsingfunction* table. The *efe_number* field is the name given to the association of equipment to the proper parsing procedure, e.g. LVM_PARSING_LVM.

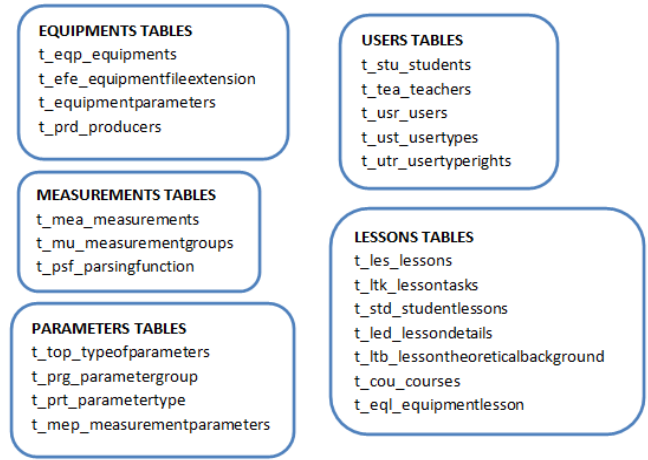

Fig. 8. The EquiLAB Database Tables

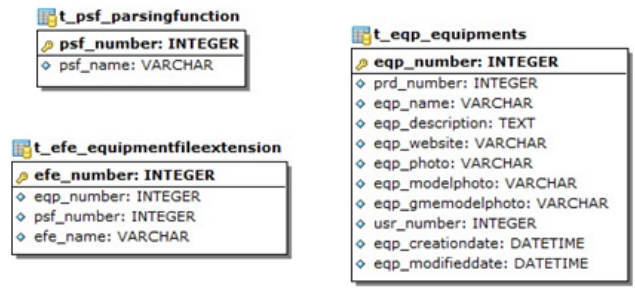

Fig. 9. Part of the EquiLAB Database Model.

V. CONCLUSIONS

The work in an engineering laboratory requires software instruments that integrate both technical and educational aspects. One needs support for performing measurement experiments with a large variety of systems, for integrating students' data, but also for describing equipment in a well structured and accessible way, and for creating attractive lessons, based on Web 2.0 capabilities. This paper presented an implementation created to respond to these challenges, with validation for an ensemble based on thermocouple signal acquisition. The use of LabVIEW in conjunction with the measuring instruments and with our integration software allows one to reuse the parsing procedures for any acquisition performed with the same workbench. The configuration only

**Annex 1- Example of *.lvm* File**

```
LabVIEW Measurement
Writer_Version      2
Reader_Version      2
Separator    Tab
Decimal_Separator   ,
Multi_Headings      No
X_Columns           One
Time_Pref           Absolute
Operator     Profesor
Date         2013/02/06
Time         17:49:40,8399038314819335937
***End_of_Header***

Notes     X values guaranteed valid only for Channel 0

Channels   3
Samples    1          1          1
Date       2013/02/06    2013/02/06    2013/02/06

Time       17:49:40,8399038314819335937
           17:49:40,8399038314819335937
           17:49:40,8399038314819335937
X_Dimension       Time       Time       Time
X0        0,0000000000000000E+0       0,0000000000000000E+0
          0,0000000000000000E+0
Delta_X     1,000000   1,000000   1,000000
***End_of_Header***
X_Value   Channel 0  Channel 1  Channel 2  Comment
0,000000   23,400000  23,400000  23,600000
0,531250   23,400000  23,400000  23,600000
1,531250   23,400000  23,400000  23,600000
2,531250   23,400000  23,400000  23,600000
.....
36,531250  23,600000  23,600000  23,799999
37,531250  23,600000  23,600000  23,799999
38,531250  23,600000  23,600000  23,799999
39,531250  23,600000  23,600000  23,799999
.....
49,531250  23,799999  23,799999  24,000000
50,531250  23,799999  23,799999  24,000000
51,531250  23,799999  23,799999  24,000000
52,531250  23,799999  23,799999  24,000000
.....
61,531250  24,000000  24,000000  24,200001
62,531250  24,000000  24,000000  24,200001
63,531250  24,000000  24,000000  24,200001
64,531250  24,000000  24,000000  24,200001
```